# Modelling mass distribution of the Milky Way galaxy using Gaia's billion-star map

Enbang Li

School of Physics, Faculty of Engineering and Information Sciences, University of Wollongong, NSW 2522, Australia

enbang@uow.edu.au

**The Milky Way galaxy is a typical spiral galaxy which consists of a black hole in its centre, a barred bulge and a disk which contains spiral arms[1, 2]. The complex structure of the Galaxy makes it extremely difficult and challenging to model its mass distribution, particularly for the Galactic disk which plays the most important role in the dynamics and evolution of the Galaxy[3]. Conventionally an axisymmetric disk model with an exponential brightness distribution and a constant mass-to-light ratio is assumed for the Galactic disk[4]. In order to generate a flat rotation curve, a dark halo has also to be included[5-7]. Here, by using the recently released Gaia billion-star map[8], we propose a Galactic disk mass distribution model which is based on the star density distribution rather than the brightness and mass-to-light ratio. The model is characterized by two parameters, a bulge radius and a characteristic length. Using the mass distribution model and solving the Poisson equation of the Galaxy, we obtain a flat rotation curve which reproduces the key observed features with no need for a dark halo.**

Modelling the mass distribution in the Milky Way galaxy is of utmost importance not only for correctly estimating the total mass of the Galaxy, but also for explaining the observed flat rotation curve which has become one of the greatest unsolved mysteries in astrophysics[9,10], and therefore has attracted numerous studies devoted to this field[3]. Historically the surface mass density of a spiral galaxy is modelled by an exponential disk as first proposed by Freeman and a constant mass-to-light ratio across the disk radius[4,11]. Clearly, the validity of this approach strongly depends on whether or not the two conditions are simultaneously satisfied. It is well known that the luminosity of a star is dependent not only on its mass, but also on other intrinsic properties including its age and chemical composition, and also on the observing bands[12,13]. As the observed results of various spiral galaxies show, the exponential disk models only apply for certain regions along the galactic radius, depending on the type of the galaxy under question[4,12]. The deviation from an exponential surface density would certainly cause systematic uncertainties in all calculations that would follow. Furthermore, the constant mass-to-light ratio assumption could be more problematic since the mass-to-light ratio can vary significantly from the galactic centre, to the innermost regions and to the outskirts of a spiral galaxy[14].

In order to avoid the issues associated with the uncertainties on surface brightness and mass-to-light ratio, here we propose to utilize the star density of the Milky Way galaxy to model its stellar mass density distribution in the Galactic disk. This is made possible by the recent data release (DR1) of the European Space Agency (ESA) Gaia satellite. The ESA Gaia mission has obtained highly accurate parallaxes and proper motions of over one billion sources brighter than G=20.7[8]. The 1.14 billion stars detected by Gaia satellite is an unprecedented sampling of the total number of stars in the Galaxy which is estimated as 200 billions[2].

As have been done previously, we discompose the mass of the Galaxy into two components: (1) a central bulge with a black hole in the centre; and (2) a Galactic disk which consists of stellar objects and interstellar medium (ISM), mainly gas.

Since the main focus of this study is the mass distribution of the Galactic disk and its influence on the overall rotation curve, here we combine the black hole and the central bulge together in the calculations of the rotational velocities, instead of as separate mass objects. This treatment will not alter the overall rotation curve significantly especially in the outskirts of the Galaxy.

Hydrogen is the most abundant element in space and can also serve as a tracer for gas and other interstellar medium (ISM) components in the Universe. The 21-cm emission line of neutral atomic hydrogen (HI) provides us with a reliable tool for quantitatively mapping the distribution of ISM in the Galactic disk. The observational results of several large-scale 21-cm line surveys are currently available with improved sensitivity, dynamic range and data quality[15,16]. Here we utilize the published HI survey data to infer the ISM mass distribution in the Galactic disk.

For the central bulge, we adopt the well-known de Vaucouleurs profile defined as[17,18]

$$\Sigma_b(r) = \Sigma_{b0} exp[-(r/r_0)^{1/4}], \qquad (1)$$

where, $\Sigma_{b0}$ is the central surface mass density of the bulge, $r$ is the Galactic radius, $r_0$ is the characteristic radius of the bulge and is determined by the effective radius $r_e$ at which the mass density of the bulge drops to half of its maximum. $r_0$ and $r_e$ satisfy a relation given by $r_e = b^n r_0$, where $b$ is a solution of equation $\gamma(b, 8) = \Gamma(8)/2$ with $\gamma$ and $\Gamma$ are respectively the incomplete and complete gamma functions. In order to determine the bulge parameters, $\Sigma_{b0}$ and $r_e$, we calculate the rotation velocity contribution from the bulge using a method as detailed below and fit it to the central part (0 to 2 kpc) of the observed rotation curve (data taken from ref.7). It has been found that taking $r_e = 0.75\ kpc$ and $\Sigma_{b0} = 5.12 \pm 0.37 \times 10^6 M_\odot/pc^2$ gives a best curve fit to the observed rotation velocities from the Galactic centre to $r = 2\ kpc$. The uncertainty for the central surface mass density is derived from the average error of the observed rotation velocities used for the curve fitting process.

The observational results of several large-scale 21-cm line surveys on the gas distributions in the Galaxy show that there two areas with high concentrations of HI and H2 gas at $r_{g1} = 5\ kpc$ and $r_{g2} = 12\ kpc$ respectively[19]. We therefore model the gas mass distribution by Gaussian distributions as

$$\Sigma_g(r) = \Sigma_{g01} \exp\{-[(r-r_{g1})/w_{g1}]^2\} + \Sigma_{g02} \exp\{-[(r-r_{g2})/w_{g2}]^2\}, \qquad (2)$$

where, $\Sigma_{g01}$ and $\Sigma_{g02}$ are the peak surface mass densities, and $w_{g1}$ and $w_{g2}$ are the widths of the Gaussian distributions.

The main focus of this work is on the stellar mass distribution in the Galactic disk which plays the most critical role in determining the profile of the rotation curve and total mass of the Galaxy. A galactic distribution of all Gaia DR1 sources on the sky is illustrated in Fig.1**a**. In order to obtain a two dimensional surface mass density from the star map, we first project the star distribution in the galactic coordinates into the Galactic plane by summing up the number of stars contained in a latitude angle range between 25 degrees and -25 degrees for each longitude angle. Then we resample the star numbers in a 2-degree bin for every 10 degree longitude angle to achieve a star number distribution, as shown in Fig.1**b**. Since an axisymmetric disk model is used here, the data from longitude angles from 0 to 180 degrees and that from 180 to 360 degrees are then averaged, and the error bars shown in the figure are from the averaging process. It can be seen that large fluctuations appear in a region covering 0 to 60 degree longitude angles. This is obviously caused by the presence of dust clouds in these areas that absorb the starlight and produce the extinction along the line of sight in the Galactic disk, as clearly shown in the star map. The star density data in a longitude angle range from 70 to 180 degrees are smoothly distributed and hence are used for determining the stellar mass distribution in the Galactic disk as described below.

From the Gaia star number distribution, we adopt a surface stellar mass density for the Galactic disk as given by

$$\Sigma_s(r) = \frac{\Sigma_{s0} l_c}{\sqrt{(r-r_b)^2 + l_c^2}} , \qquad (3)$$

where, $\Sigma_{s0}$ is the maximum surface mass density of the stellar disk, $r_b$ is the radius of the central bulge, and $l_c$ is a characteristic length of the stellar disk.

By applying the same method as that used in calculating the Gaia star number density to the bulge model and the stellar disk model as described by Eqs.(1) and (3) respectively, we obtain a modelled star number distribution as shown in Fig.1**b**. A best-curve fitting process gives: $r_b = 2.0 \ kpc$, $l_c = 2.5 \ kp$. It can be seen that the modelled curve agrees well with the data from the Gaia star map, particularly in a longitude angle range from 70 to 180 degrees.

We assume that the Gaia star map is a sample of all stars in the Galaxy and the stellar mass distribution in the Galactic disk is proportional to the star mass distribution. Therefore, the only free parameter in the surface mass density expression is the maximum surface mass density of the stellar disk, $\Sigma_{s0}$, which can be determined by fitting the observed rotation curve as described below.

Taking Eq.(1) as the mass distribution for the central bulge and assuming a spherical shape, one can calculate the rotation curve inside and outside the bulge by using[18]

$$V_b^2(r) = \frac{G\Sigma_{b0}}{r_0} \int_{m=0}^{r} \left[ \int_{\kappa=m}^{\infty} \frac{e^{-(\frac{\kappa}{r_0})^{1/4}} (\kappa/r_0)^{-3/4}}{\sqrt{\kappa^2 - m^2}} d\kappa \right] \frac{m^2}{r} dm, \qquad (4)$$

where, $G$ is the gravitational constant.

For the Galactic disk, if we treat it as a thin disk with a surface mass density as described by Eq.(3), then the Poisson's equation for the Galaxy, $\nabla^2\Phi = 4\pi G\rho$, can be replaced by Laplace's equation in cylindrical coordinates with a thin disk mass density in the Galactic plane as the boundary. The circular velocity can be expressed as[20]

$$V^2(r) = 2\pi r G \int_0^\infty \left[\int_0^\infty \Sigma(r')J_0(kr')r'dr'\right]J_1(kr)kdk, \tag{5}$$

where, $J_0(x)$ and $J_1(x)$ are the Bessel functions of the first kind with order zero and one, respectively.

Using Eqs.(4) and (5), and the mass density models for the central bugle, the stellar disk and gas disk, we can calculate the circular velocity components, $V_b$, $V_s$ and $V_g$ individually, and then the combined circular velocity to produce a rotation curve of the Galaxy by utilizing

$$V = \sqrt{V_b^2 + V_s^2 + V_g^2 - V_l^2}, \tag{6}$$

where, $V_l$ represents an effect produced by the low mass density areas (as described below) to the circular velocity and the minus sign indicates the negative contribution to the gravitational potential.

We determine the maximum surface mass density of the stellar disk, $\Sigma_{s0}$, by fitting the calculated rotation velocities to the observed rotation curve between 2.5 kpc and 8.3 kpc (the solar system location used in this study) as $\Sigma_{s0} = 611\pm44\ M_\odot/pc^2$. The uncertainty is determined by using the average error of the observed rotation velocities from 2.5 to 8.3 kpc. All parameters used in the calculations are summarized in Table 1.

We also consider the influence of the low density areas between spiral arms, as marked with "A" and "B" in Fig.2**a**. We adopt that the locations are at $r_{l1} = 3.0\ kpc$ and $r_{l2} = 9.5\ kpc\ kpc$ respectively[21]. As there are very limited observation data available to refer to, we make an assumption that at the two locations the mass densities drop with a Gaussian profile and by 50% of their local values with $\pm20\%$ uncertainties. Essentially these mass densities are the modifications to the stellar disk mass density and should be subtracted from the disk mass density. In Table 1 we equivalently list them as separate components which should be subtracted from the disk mass.

The calculated rotation curves together with the observed data are shown in Fig.2**b**. It can be seen clearly that the combined rotation curve is almost flat across the 25 kpc galactic radius considered in this study, and precisely fits the observed rotation velocities in both the innermost regions and the outskirts of the Galaxy. In addition to the overall profile agreement, the velocity drops in the rotation curve at $r = 3\ kpc$ and $r = 10\ kpc$ regions are reproduced and predicted by the mass distribution models considered here. It has been found that the gas distribution in the 12 kpc area largely contributes to the velocity drop in the rotation curve around an area near $r = 10\ kpc$. The reason is that the ring shaped gas distribution produces negative gravitational potential to the areas contained inside and close to the ring.

From the mass distributions, we also calculate the masses of the central bugle (with the black hole included), the stellar disk and the gas disk, and the results are presented in Table 1. The calculated bulge mass of $1.69 \pm 0.12 \times 10^{10} M_\odot$ lies in between bulge masses of $1.3 \times 10^{11} M_\odot$ determined by COBE NIR luminosity[22] and $2.0 \times 10^{11} M_\odot$ by using red clump giants (RCGs) [23]. It is also precisely matches a bulge stellar mass of $1.63 \times 10^{11} M_\odot$ of the M90 model in Ref.24.

The mass of the gas disk is calculate as $8.43 \pm 0.84 \times 10^9 M_\odot$, which is smaller than a recently published value of $12.3 \times 10^9 M_\odot$ within a 60 kpc radius[16].

The stellar mass of the Galaxy contained in a 25 kpc radius is estimated as $2.32 \pm 0.24 \times 10^{11} M_\odot$. The total mass of the Galaxy contained in the 25 kpc radius, including all components from the central bulge (with the black hole), stellar disk and ISM disk, is calculated as $2.57 \pm 0.23 \times 10^{11} M_\odot$. This value is close to the Galactic mass of $2.1 \times 10^{11} M_\odot$ (within 25 kpc) determined by using a kinematic method[25], and to the expected baryonic mass of $2.4 \times 10^{11} M_\odot$ from other previous studies[26,27]. This Galactic mass should also be reasonable when we adopt that: (1) there are approximately 200 billion stars in the Galaxy; and (2) the average mass of all stars in the Galaxy is equal to the solar mass, implying that the star mass in the Galaxy is in the $2.00 \times 10^{11} M_\odot$ range.

In summary, the stellar mass model derived from the Gaia star map removes the uncertainties and systematics that could be caused by those on the surface brightness distribution and the constant mass-to-light ratio assumption, and ensure more reliable modelling of the Galaxy. It should be stressed here that with the Galactic disk model proposed in this study, a flat rotation curve can be generated without using dark matter halos. The model described by Eq.(3) has also been tested and applied to other spiral galaxies. Although the detailed studies on these galaxies are out of the scope of this work, as an example, we show in Fig. 3 a modelled rotation curve for the Triangulum galaxy (M33). As M33 is a bulge-free spiral galaxy, we set $r_b = 0$ in Eq.(3), and find that $l_c = 1.5\ kpc$ provides us with a best curve fit. It can be seen from Fig.3 that the modelled rotation curve perfectly matches the observed data (taken from ref.28). Once again no dark halo is used in the M33 case.

**Table 1 Galactic parameters and estimates**

| Component | Parameter | Value |
|---|---|---|
| Bulge (with a black hole) | Half maximum radius<br>Central mass density<br>Mass | $r_e = 0.75\ kpc$<br>$\Sigma_{b0} = 5.12 \pm 0.37 \times 10^6 M_\odot/pc^2$<br>$M_b = 1.69 \pm 0.12 \times 10^{10} M_\odot$ |
| Stellar disk | Bulge radius<br>Maximum surface density<br>Characteristic length<br>Mass within a 25 kpc radius | $r_b = 2.0\ kpc$<br>$\Sigma_{s0} = 611 \pm 44 M_\odot/pc^2$<br>$l_c = 2.5\ kpc$<br>$M_s = 2.32 \pm 0.24 \times 10^{11} M_\odot$ |
| Gas (HI and H2) | Location 1<br>Location 2<br>Gaussian width<br>Gaussian width<br>Peak surface density<br>Peak surface density<br>Mass | $r_{g1} = 5.0\ kpc$<br>$r_{g2} = 12.0\ kpc$<br>$w_{g2} = 4.0\ kpc$<br>$w_{g2} = 4.0\ kpc$<br>$\Sigma_{g01} = 11.0 M_\odot/pc^2$<br>$\Sigma_{g02} = 11.0 M_\odot/pc^2$<br>$M_g = 8.43 \pm 0.84 \times 10^9 M_\odot$ |
| Low density areas | First location<br>Second location<br>Total width at $1/e^2$<br>Mass | $r_{l1} = 3\ kpc$<br>$r_{l2} = 9.5\ kpc$<br>$w = 4.0\ kpc$<br>$M_l = 2.93 \pm 0.57 \times 10^{10} M_\odot$ |
| Total Galactic mass | Within 25 kpc radius | $M_{total} = 2.57 \pm 0.23 \times 10^{11} M_\odot$ |

**References**

1. Sparke, L. S. & Gallagher, J. S. *Galaxies in the universe : an introduction*. Second ed edn, (Cambridge University Press, 2007).

2. Freedman, R. A. & Kaufmann, W. J. *Universe*. 6th ed edn, (W.H. Freeman, 2002).

3. Bland-Hawthorn, J. & Gerhard, O. The Galaxy in Context: Structural, Kinematic, and Integrated Properties. *Annual Review of Astronomy and Astrophysics, Vol 54* **54**, 529-+, doi:10.1146/annurev-astro-081915-023441 (2016).

4. Freeman, K. C. ON DISKS OF SPIRAL AND S0-GALAXIES. *Astrophysical Journal* **160**, 811-&, doi:10.1086/150474 (1970).

5. Iocco, F., Pato, M. & Bertone, G. Evidence for dark matter in the inner Milky Way. *Nature Physics* **11**, 245-248, doi:10.1038/nphys3237 (2015).

6. Alcock, C. The dark halo of the Milky Way. *Science* **287**, 74-79, doi:10.1126/science.287.5450.74 (2000).

7. Sofue, Y. Grand Rotation Curve and Dark-Matter Halo in the Milky Way Galaxy. *Publications of the Astronomical Society of Japan* **64**, doi:10.1093/pasj/64.4.75 (2012).

8. Gaia Collaboration, Gaia Data Release 1. Summary of the astrometric, photometric, and survey properties, *Astronomy & Astrophysics*, (2016), doi: http://dx.doi.org/10.1051/0004-6361/201629512

9. Rubin, V. C. The rotation of spiral galaxies. *Science* **220**, 1339-1344, doi:10.1126/science.220.4604.1339 (1983).

10. Sofue, Y. & Rubin, V. Rotation curves of spiral galaxies. *Annual Review of Astronomy and Astrophysics* **39**, 137-174, doi:10.1146/annurev.astro.39.1.137 (2001).

11. Fritz, T. K. *et al.* THE NUCLEAR CLUSTER OF THE MILKY WAY: TOTAL MASS AND LUMINOSITY. *Astrophysical Journal* **821**, doi:10.3847/0004-637x/821/1/44 (2016).

12. McGaugh, S. S. The baryonic Tully-Fisher relation of galaxies with extended rotation curves and the stellar mass of rotating galaxies. *Astrophysical Journal* **632**, 859-871, doi:10.1086/432968 (2005).

13. Courteau, S. *et al.* Galaxy masses. *Reviews of Modern Physics* **86**, 47-119, doi:10.1103/RevModPhys.86.47 (2014).

14. Portinari, L. & Salucci, P. The structure of spiral galaxies: radial profiles in stellar mass-to-light ratio and the dark matter distribution. *Astronomy & Astrophysics* **521**, doi:10.1051/0004-6361/200811444 (2010).

15. Levine, E. S., Blitz, L. & Heiles, C. The spiral structure of the outer Milky Way in hydrogen. *Science* **312**, 1773-1777, doi:10.1126/science.1128455 (2006).

16. Kalberla, P. M. W. & Kerp, J. The HI Distribution of the Milky Way. *Annual Review of Astronomy and Astrophysics, Vol 47* **47**, 27-61, doi:10.1146/annurev-astro-082708-101823 (2009).

17. De vaucouleurs, G. PHOTOELECTRIC PHOTOMETRY OF THE ANDROMEDA NEBULA IN THE U, B, V SYSTEM. *Astrophysical Journal* **128**, 465-488, doi:10.1086/146564 (1958).

18. Noordermeer, E. The rotation curves of flattened Sersic bulges. *Monthly Notices of the Royal Astronomical Society* **385**, 1359-1364, doi:10.1111/j.1365-2966.2008.12837.x (2008).

19. Wolfire, M. G., McKee, C. F., Hollenbach, D. & Tielens, A. Neutral atomic phases of the interstellar medium in the Galaxy. *Astrophysical Journal* **587**, 278-311, doi:10.1086/368016 (2003).

20. Binney, J. & Tremaine, S. *Galactic dynamics*. (Princeton University Press, 1987).

21. Sofue, Y., Honma, M. & Omodaka, T. Unified Rotation Curve of the Galaxy - Decomposition into de Vaucouleurs Bulge, Disk, Dark Halo, and the 9-kpc Rotation Dip. *Publications of the Astronomical Society of Japan* **61**, 227-236, doi:10.1093/pasj/61.2.227 (2009).

22. Dwek, E. *et al.* MORPHOLOGY, NEAR-INFRARED LUMINOSITY, AND MASS OF THE GALACTIC BULGE FROM COBE DIRBE OBSERVATIONS. *Astrophysical Journal* **445**, 716-730, doi:10.1086/175734 (1995).

23. Valenti, E. *et al.* Stellar density profile and mass of the Milky Way bulge from VVV data. *Astronomy & Astrophysics* **587**, doi:10.1051/0004-6361/201527500 (2016).

24. Portail, M., Wegg, C., Gerhard, O. & Martinez-Vapuesta, I. Made-to-measure models of the Galactic box/peanut bulge: stellar and total mass in the bulge region. *Monthly Notices of the Royal Astronomical Society* **448**, 713-731, doi:10.1093/mnras/stv058 (2015).

25. Kafle, P. R., Sharma, S., Lewis, G. F. & Bland-Hawthorn, J. KINEMATICS OF THE STELLAR HALO AND THE MASS DISTRIBUTION OF THE MILKY WAY USING BLUE HORIZONTAL BRANCH STARS. *Astrophysical Journal* **761**, doi:10.1088/0004-637x/761/2/98 (2012).

26. Bland-Hawthorn, J. Warm gas accretion onto the Galaxy, in *254th Symposium of the International-Astronomical-Union.* 241-254 (2009).

27. Smith, M. C. *et al.* The RAVE survey: constraining the local Galactic escape speed. *Monthly Notices of the Royal Astronomical Society* **379**, 755-772, doi:10.1111/j.1365-2966.2007.11964.x (2007).

28. Sofue, Y. Rotation curve decomposition for size-mass relations of bulge, disk, and dark halo components in spiral galaxies. *Publications of the Astronomical Society of Japan* **68**, doi:10.1093/pasj/psv103 (2016).

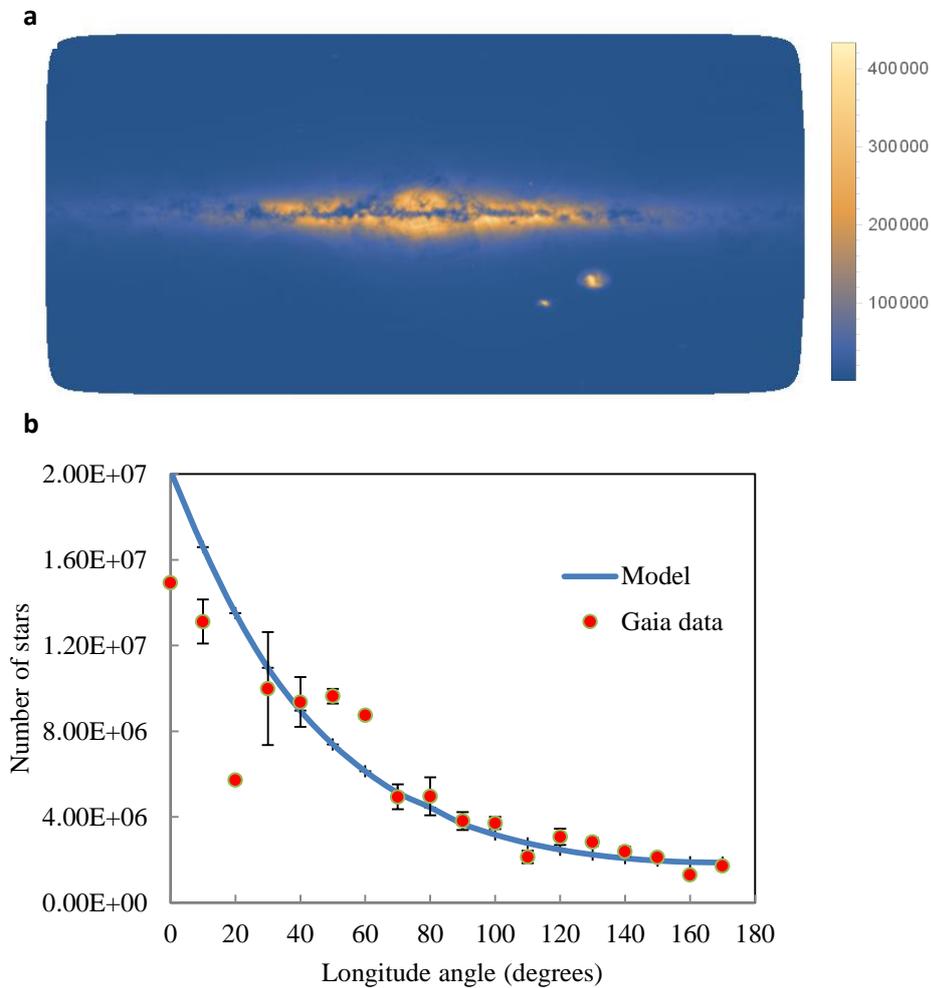

**Figure 1 | Gaia star map and star density distribution.**

**a**. A Milky Way's star map constructed from the first catalogue of more than 1.14 billion stars detected by ESA's Gaia satellite. **b**. Star numbers contained in a 2-degree bin at different longitude angles in the Galactic coordinate system. At each longitude angle, star numbers are summed up between ±25 degrees latitude angles to ensure that only stars in the galactic disk are counted in data processing. The solid line is from a mass density model proposed in this work and used in calculating the rotation curve of the Milky Way galaxy.

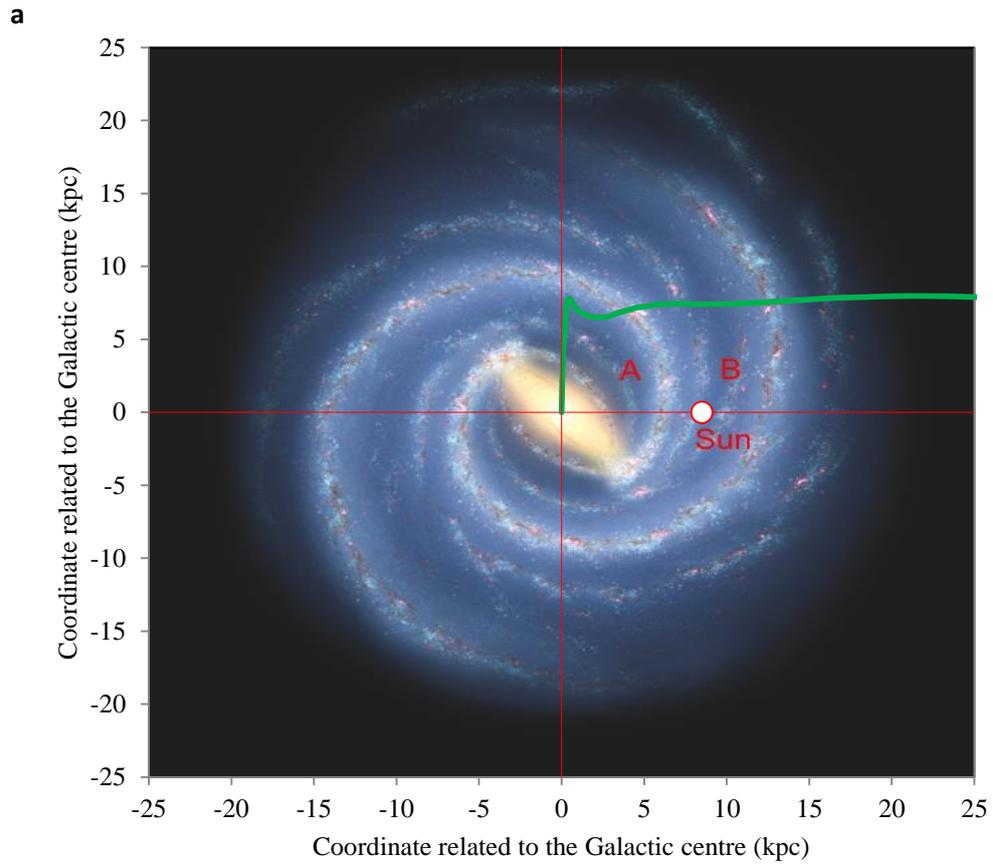

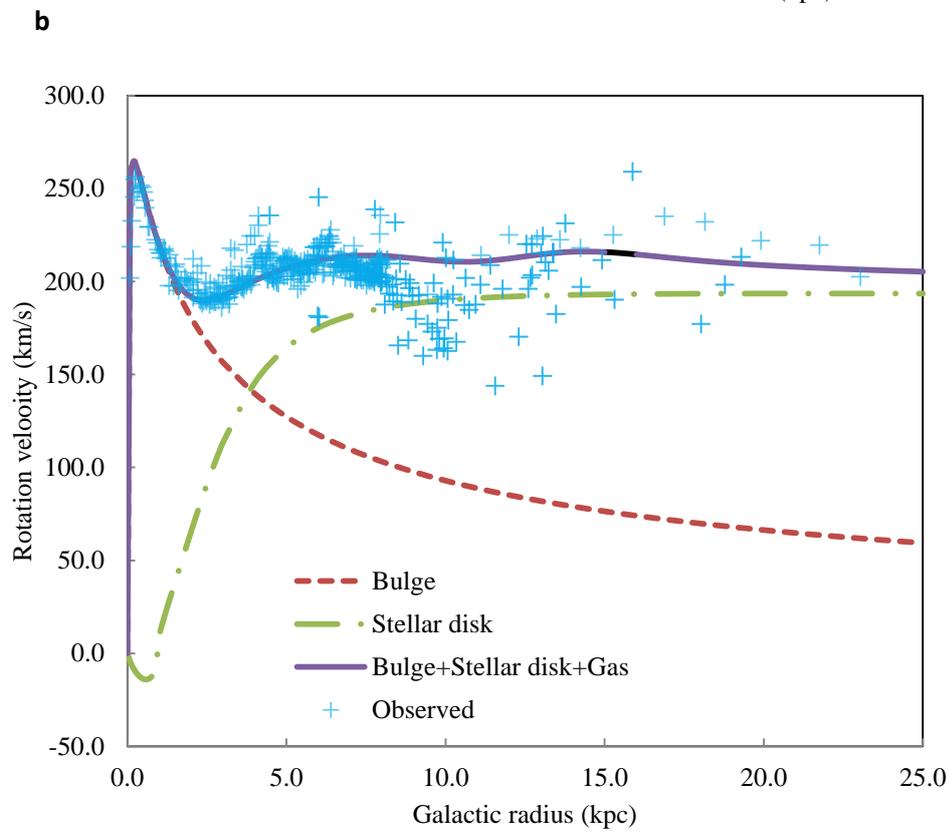

**Figure 2 | The spiral structure of the Milky Way galaxy and its rotation curves.**

**a**. Coordinate systems overlayed on an illustration image of the Milky Way galaxy (image credit: NASA). The solar system locates at $R_\odot = 8.3\ kpc$ and is marked as a white dot in the coordinate system. The green line is an observed rotation curve (data taken from ref.7) showing the flat feature across a Galactic radius up to 25 kpc. The Milky Way's complex structure which consists of a central bar-bulge and spiral arms in the galactic disk is clearly shown. Two areas marked with A and B are the two low star density areas which cause the fluctuations in the rotation curve. **b**. Rotation curve of the Milky Way galaxy calculated by using the mass model proposed in this work. The red dotted line represents the velocity component generated by the central bulge; the green dot-dashed line is the contribution from the stellar disk described by the stellar mass distribution inferred from the Gaia star map. The solid line is a combined rotation curve where the bulge, stellar disk and interstellar media (ISM) as well as the low star density areas, are considered. The blue crosses are the measured rotations velocities (ref.7). It is obvious that not only the flat feature but also the low velocity areas at $r = 3\ kpc$ and $r = 10\ kpc$ have been precisely reproduced. No dark matter halo is used in this work.

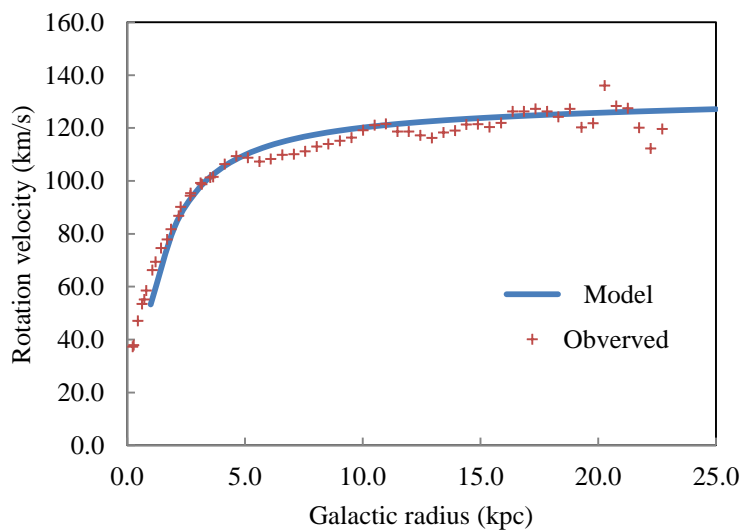

**Figure 3| Rotation curve of the Triangulum galaxy (M33).** The modelled rotation curve is obtained by using Eq. (5) while setting $r_b = 0$ (no bulge) and $l_c = 1.75\ kpc$ in Eq. (3). The observed data are taken from ref. 28.